\documentclass[sigconf,nonacm]{acmart}
\AtBeginDocument{%
  }
\usepackage{xspace}

\newcommand{\system}{\textsc{PalmGazer}\xspace}

\copyrightyear{2018}
\acmYear{2018}
\acmDOI{XXXXXXX.XXXXXXX}

\acmConference[Conference acronym 'XX]{Make sure to enter the correct
  conference title from your rights confirmation emai}{June 03--05,
  2018}{Woodstock, NY}
\acmPrice{15.00}
\acmISBN{978-1-4503-XXXX-X/18/06}

\begin{document}

\title{\system: Unimanual Eye-hand Menus in Augmented Reality}

\author{Ken Pfeuffer}
\affiliation{%
  \institution{Aarhus University}
  \city{Aarhus}
  \country{Denmark}}
   \email{ken@cs.au.dk}

  \author{Jan Obernolte}
\affiliation{%
  \institution{LMU}
    \city{Munich}
  \country{Germany}}  
  \email{jan.obernolte@campus.lmu.de}

  \author{Felix Dietz}
\affiliation{%
  \institution{Bundeswehr University Munich}
    \city{Munich}
  \country{Germany}}  
  \email{felix.dietz@unibw.de}

  \author{Ville Mäkelä}
\affiliation{%
  \institution{University of Waterloo}
    \city{Waterloo}
  \country{Canada}}  
  \email{ville.makela@uwaterloo.ca}
  
  \author{Ludwig Sidenmark}
\affiliation{%
  \institution{University of Toronto}
    \city{Toronto}
  \country{Canada}}  
  \email{lsidenmark@dgp.toronto.edu}

\author{Pavel Manakhov}
\affiliation{%
  \institution{Aarhus University}
  \city{Aarhus}
  \country{Denmark}}
   \email{pmanakhov@cs.au.dk}
   
\author{Minna Pakanen}
\affiliation{%
  \institution{Aarhus University}
  \city{Aarhus}
  \country{Denmark}}
   \email{mpakanen@cc.au.dk}

  \author{Florian Alt}
\affiliation{%
  \institution{Bundeswehr University Munich}
    \city{Munich}
  \country{Germany}}  
  \email{florian.alt@unibw.de}

\renewcommand{\shortauthors}{Pfeuffer et al.}

\begin{abstract}
How can we design the user interfaces for augmented reality (AR) so that we can interact as simple, flexible and expressive as we can with smartphones in one hand? To explore this question, we propose PalmGazer as an interaction concept integrating eye-hand interaction to establish a singlehandedly operable menu system. In particular, PalmGazer is designed to support quick and spontaneous digital commands-- such as to play a music track, check notifications or browse visual media -- through our devised   three-way interaction model: hand opening to summon the menu UI,  eye-hand input for selection of items, and dragging gesture for navigation. A key aspect is that it remains always-accessible and movable to the user, as the menu supports  meaningful hand and head based reference frames. We demonstrate the concept in practice through a prototypical personal UI with application probes, and describe technique designs specifically-tailored to the application UI.  A qualitative evaluation highlights the system's design benefits and drawbacks, e.g., that common 2D scroll and selection tasks are simple to operate, but higher degrees of freedom  may be reserved for two hands. Our work contributes interaction techniques and  design insights to expand AR's uni-manual capabilities.
\end{abstract}



\maketitle

\section{Introduction}
\begin{figure}[t]\includegraphics[width=1\columnwidth]{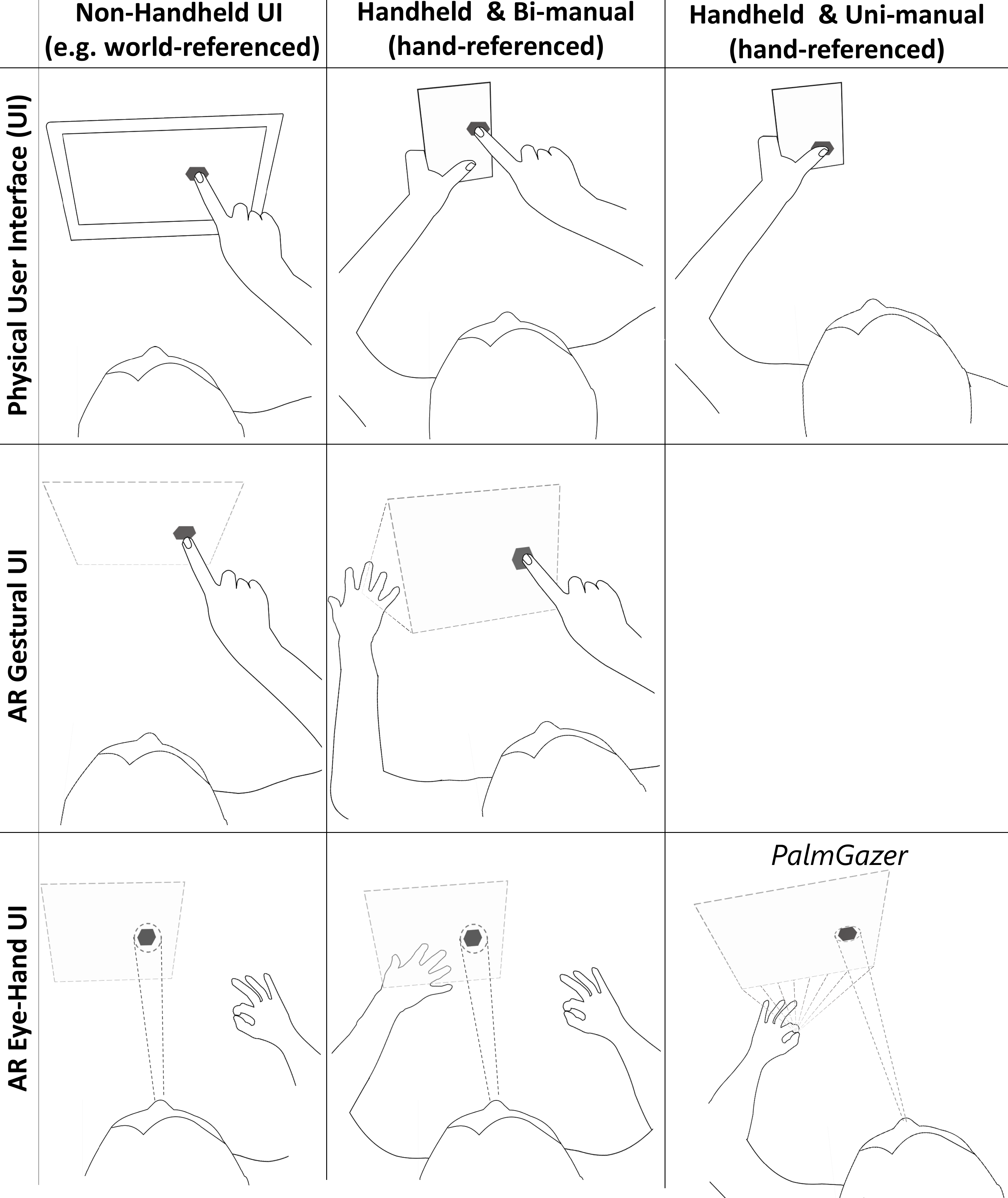} \vspace{-0.7cm}
\caption{A taxonomy for one-handed computer interaction across physical and spatial user interfaces. Everyday physical devices support the primary handheld and one-handed interaction classes for 2D menu UIs (top row). In contrast, gestural and eye-hand UIs only partially cover one-handed interaction. \system fills the gap through an eye-hand UI to facilitate one-handed  interaction for AR users.}
\label{handedness}\end{figure}

Whereas the user interface (UI) for augmented reality (AR) head-mounted devices (HMD) is maturing for stationary settings like at home or at work, less has been focused on AR UIs  tailored for aiding people in their daily life like smartphones do presently. Everyday environments are highly dynamic and confront users with a variety of information which can lead to situations where users have limited physical and attentional resources available. One of the most common types of scenarios is when just one hand is available. Consider the case with smartphones. They allow to readily grip the device to bring it into view and support a variety of actions from spontaneously playing a song to extended sessions browsing photographs, reading messages, and navigating a map --all unimanually. A phenomenon that enables this is the duality  of two  aspects: i) the device is \textit{handheld} which allows to bring the content into the ideal viewing place, and ii) the device is \textit{interactive} through few, simple gestures that cover an expressive range of UI controls.

Such basic yet expressive one-handed interactions are desirable but only partially supported in gestural AR UIs (Figure \ref{handedness}). The landscape for one-hand interaction is two-fold.
First, handheld UIs such as the Hand Menu \cite{Handmenu} support one hand performing all UI activities. They also use the non-dominant hand as reference frame for a UI. This enables users to use their human sense of proprioception to stay aware of where the UI is in relation to the body \cite{Mine97}. Taken together, this would be typically classified as a case of asymmetric bi-manual interaction according to Giuard \cite{Guiard87}.
Second, users may easily interact uni-manually when the UI is detached from the user's hand and instead is  for instance fixed to any location in the physical world. This  ties the user to a stationary place, which is useful in situations where you remain for an extended amount of time.  However, if a user's position in the world space changes, he or she must manually relocate and re-instantiate the UI, which may be time-consuming \cite{Lu22}.

A 3D interaction medium based on the eyes and hands in combination has lately received rising attention  in the Human-Computer Interaction (HCI) literature \cite{Wagner23,lystbaek2022alignment,Mutasim21,Fernandes23}.  AR HMDs (e.g., Microsoft HoloLens, Meta Quest, Magic Leap, and Apple Vision Pro)  integrate variations of this medium. 
Eye-hand interaction has so far been proposed for the same stationary cases as for gestural UIs: world-referenced UIs for one-handed tasks \cite{MRTKgazezoom} and handheld UIs for two-handed tasks  \cite{Pfeuffer17,Pfeuffer21}. 
In this work, we take a distinct direction by exploring how the two modalities can facilitate, and if yes to what extend, handheld unimanual interaction.

\textbf{\system} is a novel UI system for AR HMDs that support eye and hand tracking. \system combines  eye and hand gestures to establish a simple but sufficiently-rich interaction model to afford expressive one-handed menu operations. At its core, it covers three canonical fundamental interaction tasks through three mutually-compatible interaction techniques:

\begin{figure*}
  \includegraphics[width=\textwidth]{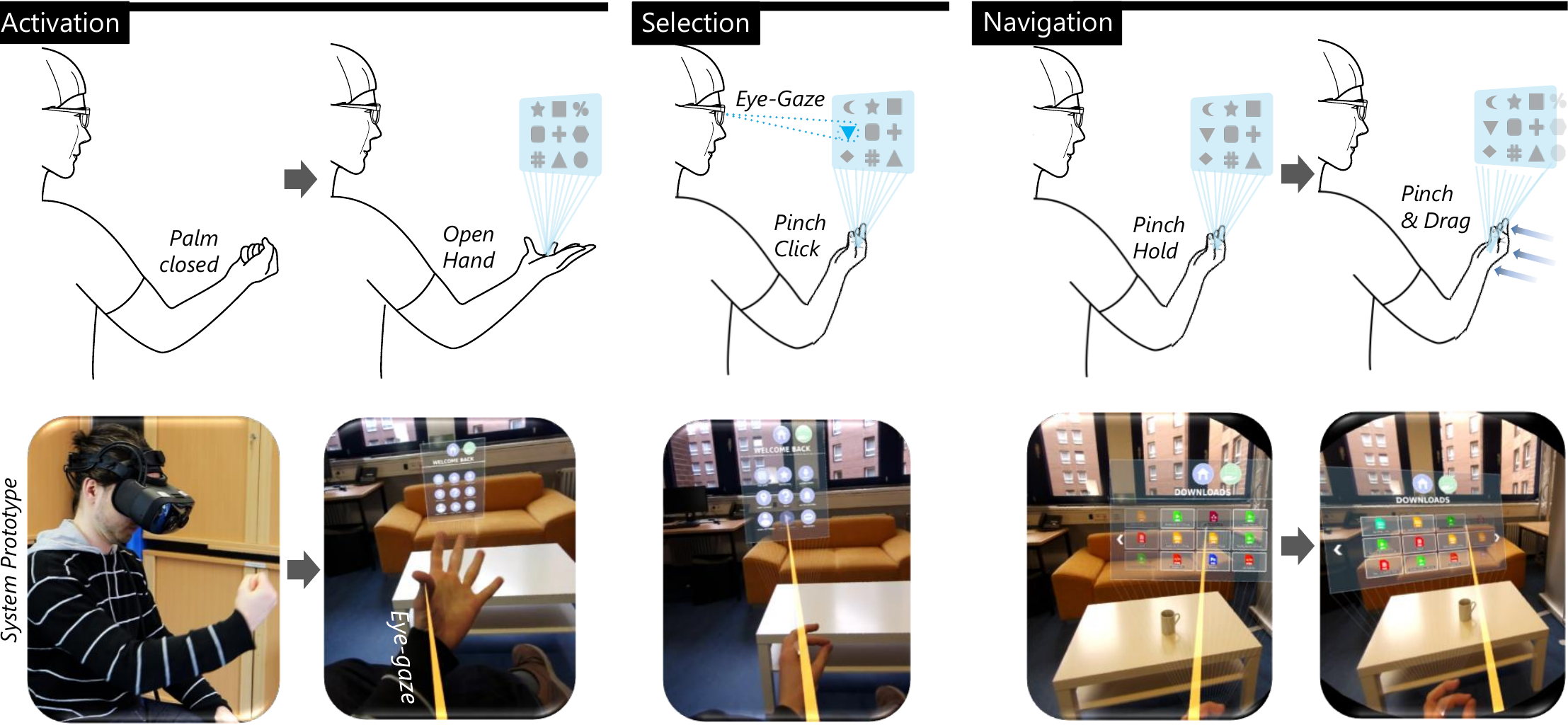}
  \vspace{-0.7cm}
  \caption{\system integrates three core techniques to form a unimanual, on-demand UI: (a) UI activation can be performed by a palm-open gesture that summons the UI above the hand, allowing on-demand access by one hand when needed. (b) Selection is accomplished through gazing at the item of interest and a pinch gesture. (c) Navigation commands via pinch-dragging involve both content transformation and UI movement.}
  \label{teaser}
\end{figure*}


\begin{itemize}
\item \textbf{UI Activation (\autoref{teaser}-left)}: It starts with the user activating the UI by a palm-open gesture that summons a hand-attached home menu. Hereafter they are free to adjust the UI position by hand to place it  mid-air as desired, as the UI remains active as long as the palm is opened. The user closes their hand to disengage the UI at any moment of time, facilitating the notion  that it is readily available and easily dismissable at will.
\item \textbf{Selection (\autoref{teaser}-center)}: Instead of then engaging a second hand, the interaction design is fully tailored to  the same hand -- via pinch gestures while the palm is opened. To select an object in the menu, eye-hand input is employed in form of gaze  for target acquisition and a quick-release pinch gesture for confirmation. This allows for compound interactions where users can rapidly dovetail palm-open, look, and pinch for a rapid one-off action.
\item \textbf{Navigation (\autoref{teaser}-right)}: Basic navigation commands such as scrolling the whole UI make the UI expressive beyond a page. Here performed by a pinch dragging gesture in the respective direction, it involves a design conflict. Moving the hand for scrolling inevitably means the handheld UI is moved in its place. We address this by a peephole-inspired UI behaviour that retains spatial relationships of content in space.

\end{itemize}
All three techniques may be effortlessly integrated into a compound system operation. 
From an input-theoretic standpoint, a first key benefit is that the three techniques establish  a coherent and simple input command language for one hand only-- but lends itself to be as expressive as two-handed or non-hand-referenced UIs. A second key benefit is the menu is always available and easily dismissable at will, allowing for spontaneous actions on the go. In light of this, our main objective is to understand what kind of features and applications are suitable for such a one-handed AR UI concept as formulated in the following research question:

\textit{How expressive can \system be, how many of the basic mobile interactions can be covered by one hand, and at which point is a task more suitable for a two-handed approach?} 

To explore these questions, our work takes  a systems-oriented approach, where we design and implement a holistic UI that includes interaction techniques and AR application probes, forming a particular balance between expressive power, ease of use, and multimodal fusion. Such a holistic approach  allows  to gain breadth-first insights into low-level task- and application-specific design issues and trade-offs, and to better understand the  high-level integration and compatibility of \system as a whole. Lastly, this system is used in an informal study, where users  get systematically experienced with simple \textit{vs.} complex interaction techniques and hand \textit{vs.} head based UI reference frames. 

We find that the holistic UI concept was, after a brief training phase, easy to use for the study participants. Interestingly, the ability to move the UI farther or closer to the eyes facilitates eye-tracking interactions as users can dynamically resize the visual target size. With regards to expressiveness, we find that all basic actions for selection and navigation are suitable, while higher degrees-of-freedom tasks become too challenging with one hand and gaze only. These findings contribute to the prior knowledge by proposing a novel approach to the class of fully one-handed interaction, and provision of a better understanding of the merits and limitations across a variety of usability factors.

Our contributions include four points. First, our design is grounded in an analysis of relevant eye-hand AR UI systems that expands the understanding of the fragmented nature of UI activation and interaction techniques in research and industry. Second, we present new interaction concepts as basis for \system, including a consistent set of selection and navigation techniques for 1D/2D/3D transformations, and the reference frames of  \textit{Above-Hand} and \textit{On-Hand} with distinct ergonomics/visibility trade-offs offered as choices for users. Third, a prototype system showcased by 6 applications to demonstrate  how expressive the input concept is, and to selectively highlight interesting design decisions to take for system-wide and application-specific actions. Lastly, we present insights from an informal evaluation of the system,  revealing the conceptual and technical merits and limitations.


%
%

\section{Related Work}

AR UIs can be put in 3D space along various frames of reference such as body-attached, hand-referenced, or references relative to the world. In this context, major challenges involve (1)  UIs in the physical world can be easily left unnoticed and need time-consuming and physical effort to move them around, (2) menus staying continuously in the field of view  obstruct the reality of the user, and (3) a button for activation can be distracting, difficult to hit, and the number of buttons on input devices is limited as well as adds an unneeded step to user interaction \cite{Lu22,Mine97}.


A related interaction concept are  adaptive UIs that  exploit contextual information of the user and the environment to implicitly and automatically place and present the digital information to the user \cite{Belo22,Cheng21}. 
A set of works investigated the use of  eye-tracking information to infer attention to a particular widget in AR, based on a concept where virtual information of the widget reacts to the user's attention, and  gradually expands the passive perception of information  \cite{Davari22,Lu20,Lu21,Pfeuffer21,piening21}. Such glanceable widgets are appropriate for simple notifications and information retrieval \cite{piening21}, but are less suitable for expressive menu interaction. As such, Lu et al. highlight the need for specific activation techniques to render the UI more expressive and avoid cluttering the UI \cite{Lu20,Lu21}, for which we provide  a solution. In principle,  context-aware and glanceable UIs can  improve the user's efficiency in accessing information, pointing to the potential of AR to advance mobile interaction \cite{Davari22}. These efforts aim to provide implicit, hands-free interaction, while we focus on explicit one-handed interaction.

\begin{figure*}[t]\includegraphics[width=1.0\linewidth]{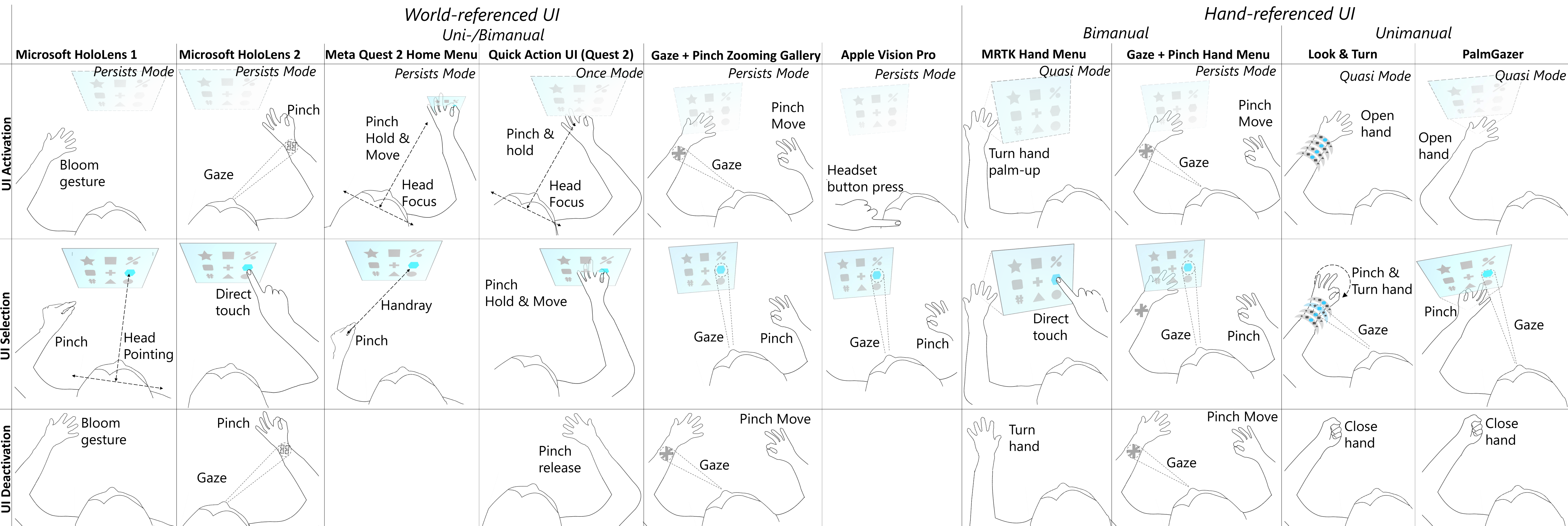}\vspace{-0.3cm}
\caption{A comparison of AR menu UIs that involve hand and eye interactions to activate and interact with virtual content.}
\label{ds}\end{figure*}

Eye-tracking as an input device is becoming established, and research efforts are increasingly  exploring integration into  menu UIs. 
For example, Ahn et al.\ have recently investigated  \emph{StickyPie} \cite{Ahn21}, where gaze  enables scale-independent marking menus that overcome overshoot errors of regular gaze-based pie menus. Yi et al.'s \emph{GazeDock}  \cite{Yi22}
employs gaze for activation of and item selection in
a view-fixed peripheral menu, finding that 4-8 items leads to the highest throughput and that the method was preferred by users over the Dwell-time and Pursuits techniques.  \emph{Radi-Eye} by Sidenmark et al.\ uses gaze- and head-interaction in a world-fixed pop-up radial UI \cite{Sidenmark21x}, where each circular level provides a submenu. In their studies, they find it affords fast and error-free interaction by nurturing on natural eye-head coordination. The concept of Gaze-Hand Alignment has been introduced by Lystbæk et al., where the mere spatial coordination between hand and gaze ray provides a fast selection mechanism for world-fixed menus in the environment\cite{lystbaek2022alignment}. These works show important milestones for gaze-enabled menu systems, which we extend by study of unimanual menus.

Eye-hand interaction  can advance the manual input capabilities of a computer user, e.g.\ when interacting with the mouse and multi-touch displays \cite{Zhai99,Pfeuffer14}. In 3D virtual environments, eye-hand interaction such as   Gaze + Pinch \cite{Pfeuffer17}, where the eyes select targets and the hands perform manipulations, has been studied in a few controlled settings with virtual and augmented reality devices for interaction with the presented 3D content. From that, a set of empirical evaluation papers have been published by researchers. These papers indicate that this type of input medium has advantageous qualities with regards to task completion times and physical effort compared to hands-only or eyes-only UIs \cite{kyto2018pinpointing,Mutasim21,Wagner23,lystbaek2022alignment,Lystbaek22text}. Pfeuffer et al. have proposed a set of concepts how eye and hand inputs can be used in combination in a 'gaze selects, hand manipulates' manner
In scientific circles, a few initial two-handed VR menus \cite{Pfeuffer17} and specific application areas were proposed \cite{Reiter22}. We extend the prior art through an exploration of how eye-hand inputs can advance one-handed menu interaction. 

In currently available AR HMDs (E.g., Meta Quest and Microsoft Holo Lens series), developers have access to gaze and hand tracking and gesture recognition software and resulting interaction techniques  (e.g., the \emph{Gaze and Commit}  technique of the Mixed Reality Tool Kit (MRTK) \cite{MRTKgazecommit}). Apple's Vision Pro is using a multimodal eye, hand, and voice UI for control of their spatial computing operating system VisionOS, showing clear potential as a new input paradigm for AR. In \system, several interaction concepts overlap with Apple's UI, such as the selection and navigation actions. We note that our work is based on a master thesis submitted long before the release of the Vision Pro \cite{jan}, and our work focuses on handheld UIs as outlined next.

\section{Characterisation of AR UIs}
User-centred UI systems must support at least a way to activate the UI as well as a way to interact with its content. A variety of modalities and techniques have been employed for those tasks in prior work and in available devices, that it is difficult to reach an overview. To shed light into this, we selected a set of closely-related systems, which means that they support hand and/or eye tracking, use primarily pinching as an intuitive and established hand tracking input \cite{Schmitz22} and offer 2D spatial windows for pointing.
Given this, we categorise and discuss the approaches  across the factors of  \textit{Mode Management}, \textit{Reference Frames}, and \textit{Handedness}, to clarify similarities and differences and lay out the design opportunity that \system tackles.

\subsection{UI Mode Management} 
 UI activation (and deactivation) techniques can be  categorised into three types of UI mode management models \cite{Hinckley06}. Current XR systems employ a controller button press and hand gesture using a \textbf{Persists} model, where the UI persists at a specific place after activation. E.g., the Meta Quest 2 offers the Quick Action menu. Here the user holds a pinch gesture, which activates a small menu with a button to open the main menu. The HoloLens 1 uses a dedicated 'bloom' hand gesture (flexing all fingers out), to be performed each time for activation/deactivation. VEIA and the Microsoft HoloLens 2 operating system offer a persistent button at the hand's forearm for activation. The user can either directly pinch at it with the other hand, or employ Gaze \& Pinch. The Apple Vision Pro, in contrast, uses a physical button on the headset. Explicit activation lends itself to suit longer sessions at a stationary place with world-referenced UIs.
The Quick Action menu itself represents a \textbf{Once}  model, as the UI opens once when holding a pinch for a set time, and closes when the pinch gesture is released. To avoid conflicting with default pinch gestures, the system employs a dwell-time during which the user  holds the gesture  at the centre of the FoV. This is  useful for infrequent, one-off actions.
In a \textbf{Quasimode} \cite{raskin2000humane} model, the UI mode is active as long as the user maintains a constant kinesthetic action, that can reduce mode errors \cite{sellen1992prevention}. 
A commonly employed method that can considered a Quasimode is UI activation  when the hand's palm is facing forward and the fingers are spread out \cite{Bowman01,Harrison12,Pfeuffer17, Zhenyi14}.  While palm-up works well for UI display, the same hand can still be used for object interaction when not held palm-up \cite{Bowman01}, representing a plausible implicit mode-switch for users.

\subsection{Reference Frames}
Many use cases include the stationary (e.g., at home, workplace, or at an exhibition), where digital content defaults in a fixed position (\textbf{world-referenced}) until people explicitly relocate or re-instantiate it. As exception, MRTK's tag-along behaviour also adapts the UI to the headset's position when it falls outside the FoV, and follows the user around. 
\textbf{Handheld} (or hand-referenced)  UIs   leverage the users' sense of proprioception to enable menus hiding at a place on the user's body that can be pulled out on demand and to allow users to deliberately hold and reposition the UI on-demand  \cite{gobeli2018effects,Mine97,Whitmire17}. This  enables  having the UI flexibly 'at hand' and avoids  potential occlusion of objects or persons in the vicinity and inconvenient automatic placements. Most closely to a mobile phone is a placement directly inside the palm and using the other hand for direct touch. However, with inside-out hand tracking, two hands can block each other in the camera's vision. As workaround, hand tracking SDKs \cite{Handmenu,ultraleap} offer UI panels next to the hand.

\subsection{Handedness} 
World-fixed UIs are commonly used with for \textbf{one-handed interaction}. But hand-attached UIs seldom enable one-handed operation, as moving the UI and pointing at it clash when done with the same hand.
To approach this, the interaction can be temporally  or spatially multiplexed. With the Quick Access menu, users can summon the UI to their hand's location where it  snaps to world space, and then use the same hand to point at a menu item. A long-pinch motion is employed as a mode switch to summon the UI, which needs to be performed each time a menu selection is desired. To afford parallel use, both activities can be spatially multiplexed. In the same hand, these specific hand postures and fine-grained finger motions can be employed to increase expressiveness \cite{Whitmire17,Gupta19,jiang22}, but can depart from the ease and popularity that pinches provide \cite{surale2019experimental,Schmitz22}.
Instead, typically a second spatial pointer is enabled via the second hand. Such a \textbf{two-handed interaction}, aligned with  Guiard's Kinematic Chain Model through  an asymmetric division of labour  \cite{buxton2008two, Guiard87}, has often been employed for 3D interaction  \cite{Handmenu, mapes1995two, Stoakley95, ultraleap,Zhenyi14,Zhang19}. 
Considering gaze input, bi-manual interaction has also shown compatibility with gaze pointing   for symmetric and asymmetric constellations \cite{Pfeuffer17,Pfeuffer20,Pfeuffer16par}. In only few cases, Gaze + Pinch has been employed for UIs on the same hand, i.e.\. in Look \& Turn or in the HoloLens 2's activation method. The UI pointing task can be offloaded to the gaze modality, offering a new option to enable simultaneous usage of holding (by hand) and interacting (by gaze and pinch), as a fundamental concept guiding \system's UI.

\subsection{\system} Based on this characterisation of 3D UI systems, \system is a unique system that encapsulates  (1)  \textit{quasimode UI mode management}, as users can rapidly activate or deactivate the UI by a hand-open or hand-close gesture; (2)  \textit{handheld reference frames}, to afford that users can place and reposition the UI at will as it is hand-attached; and (3) \textit{uni-manual interaction}, where users can simultaneously hold and interact with the UI using the same hand.

\begin{figure}[t]
    \includegraphics[width=1\columnwidth]{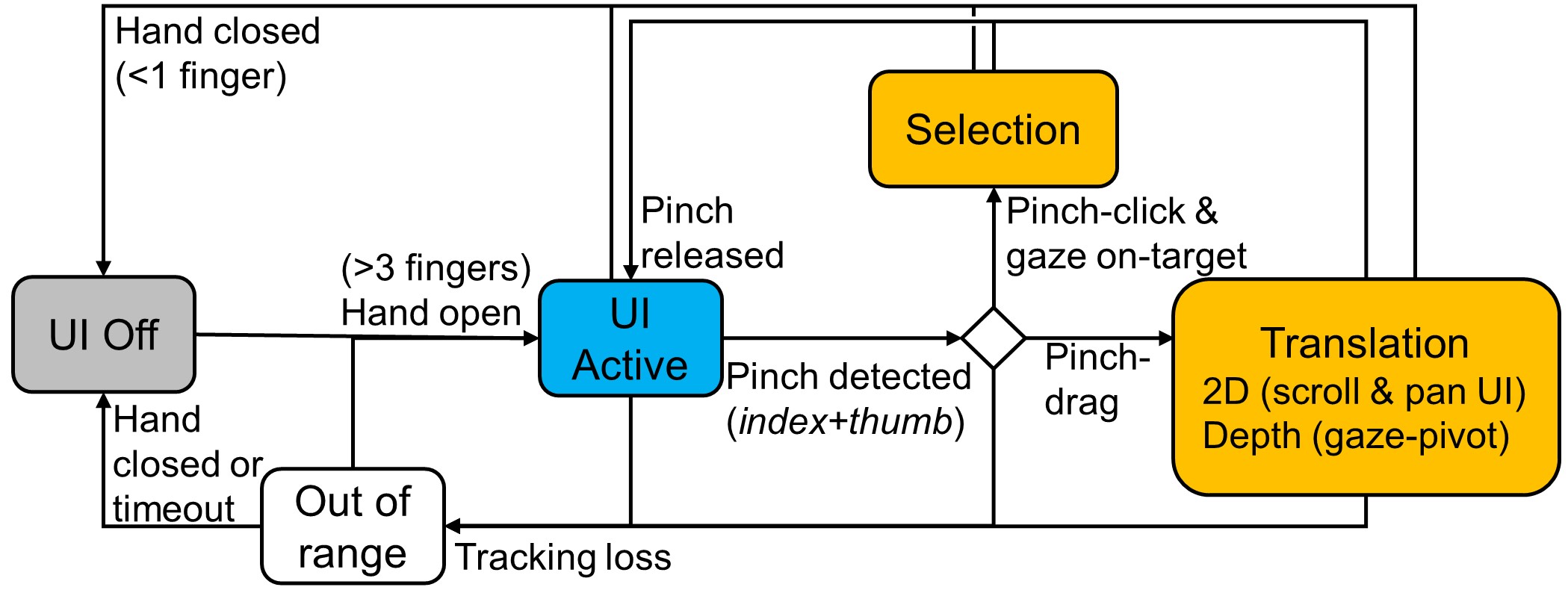} 
    \caption{Input state model of \system, based on the user's hand posture and gestures sensed by the system.}
    \label{states}
\end{figure}

\section{Design of Menu and Interaction Techniques}

%
%
An overview of the input states and transitions is shown in \autoref{states}. 
%
The UI system integrates those tasks into a state model that as receives the user's gaze and hand-tracking event information and reacts accordingly.
By default, the UI is in the \textit{UI Off} state, but opening the hand will summon it. The most recently active application is shown within the UI, and the user can interact with it. Performing a \textit{pinch-drag} gesture, e.g., for scrolling a list with overflowing elements, sets the system into the \textit{Translation} state. Issuing a \textit{pinch-click} command invokes the \textit{Selection} state action, combined with the gaze directed at a target item. The user can then simply close the hand, which returns the system to the \textit{UI Off} state.

\begin{figure}[t]\includegraphics[width=1\columnwidth]{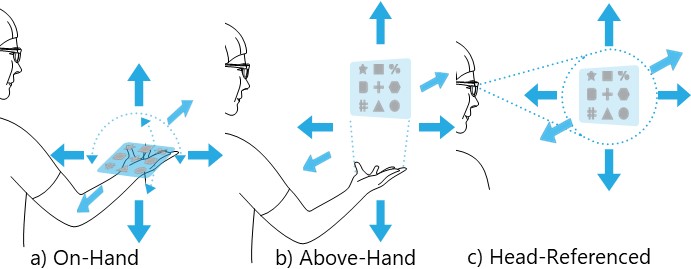} \vspace{-0.3cm}
\caption{\system supports two hand-attached UI placement concepts, one that resembles smartphone use in the hand and another where a UI 'hovers' well above the hand, compared to a typical head/world-referenced UI. }
\label{ps}\end{figure}

\subsection{Reference Frames: Hand, Above Hand, Head}
\system can utilise three reference frames for eye-hand menus. Close to a smartphone is the \textbf{On-Hand} placement (\autoref{ps}a). As UIs on the hand render overlapping hands difficult to track, they are typically not recommended  \cite{Handmenu}. Yet, via our uni-manual concept, we avoid this issue altogether. The On-Hand placement locates the UI slightly above the hand's palm centre and orients the UI according to the hand's palm. The user can shift the UI to a comfortable viewing position by moving their hand. The major benefit of its similarity to a smartphone, where the UI is always available in hand, is trading with the potential of ergonomic viewing and interaction posture. If the hand is held at the height at the field of view, prolonged use can lead to arm fatigue; if the hand is held lower (e.g., at waist level), arm fatigue is reduced, but a less ergonomic neck posture may be the case  \cite{dennerlein15, tegtmeier18}. 
 
 As an alternative, we propose the \textbf{Above-Hand} placement (\autoref{ps}b).
The 'summon' type of hand-attached UI that floats at a distance above the hand is inspired by science fiction movies where it has been intuitively leveraged for the presentation of holographic data (e.g.,\ Iron Man, Loki, or Star Wars).
We apply a similar adaptation policy to On-Hand, except that the UI is raised 30~cm above the hand, with an additional offset of 15~cm away from the user's hand in direction away from the user's head. This enables users to retain their hand in a natural posture close to their waist while maintaining a good view of the UI within their field of view.  
Lastly, we also note that while we focus on hand-attached UIs, the \system interaction technique  can be employed for \textbf{Head-Referenced} (\autoref{ps}c) or world relative UIs, for a different usage scenario where it strikes a different trade-off between UI and  real-world visibility, as it remains in place when moving.

\begin{figure}[t]\includegraphics[width=1\columnwidth]{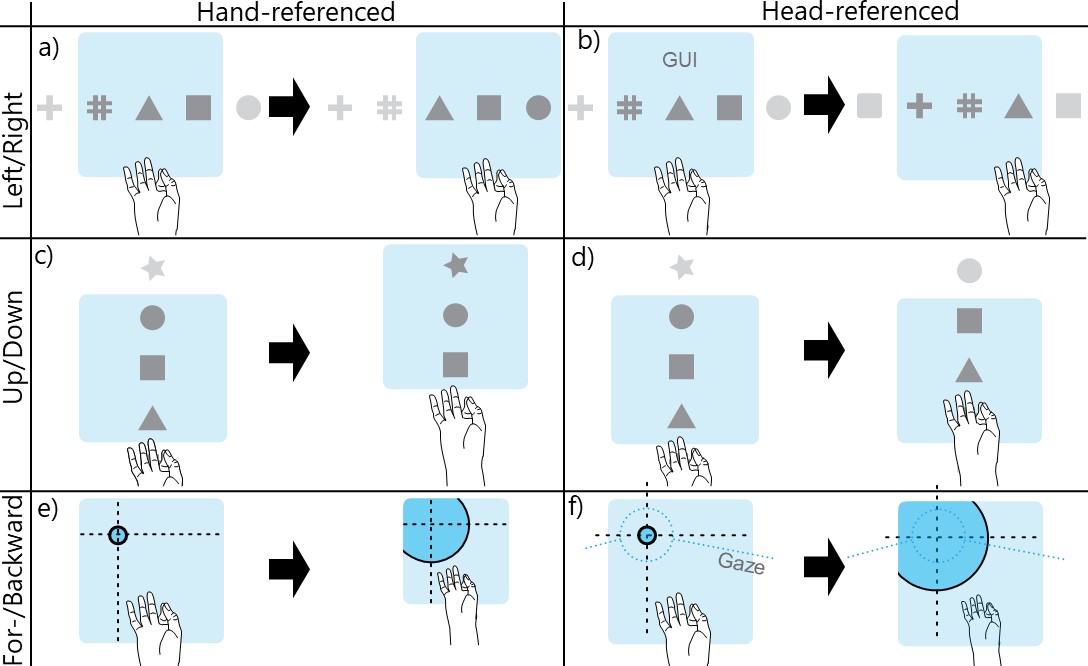} 
\vspace{-0.3cm}\caption{
As a pinch-drag gesture affects both UI and content, navigation techniques are differently handled in \system (left column) than typical head/world referenced UIs (right). }
\label{nav}\end{figure}

\subsection{2D Navigation: Static vs.\ Dynamic Peephole View}
Navigating the UI should be as simple as on a smartphone where a  swipe allows the user to see the next page or pan an image. We employ an analogous principle: when holding a pinch gesture, horizontal or vertical hand movement translates directly to a scrollable or draggable UI  (\autoref{nav}a-d). 
An interesting aspect is how directional hand motion is mapped to manipulation parameters of the UI depending on the UI placement. We consider a metaphor that the UI represents a peephole view into the virtual world allowing to retain the  spatial relationships of the virtual content \cite{Fitzmaurice93,Mehra06}. A head or world-referenced UI is more suitable for a metaphor of a \textbf{static peephole UI}, as the UI position is independent of hand motion (b, d). A hand-referenced UI is different, as when conducting a dragging gesture, the UI would move with it to sustain the spatial anchor to the hand. Here we reverse the translation direction, rendering the interaction a \textbf{dynamic peephole window}. This means the UI moves with the hand, and acts like a window to a larger virtual space behind it (\autoref{nav}a, c).

\subsection{Depth Navigation: Discrete vs.\ Continuous}

Depth navigation is a  feature in many applications of \textbf{discrete}  (e.g., a hierarchical tree structure) and \textbf{continuous translation} (e.g., zooming a map). Much like with 2D navigation described above, users can hold the pinch gesture and move their hand in the forward and backward direction from the perspective of the user. The head-referenced variant is akin to MRTK's Gaze-supported pan and zoom, where a single hand pinch gesture allows zooming a map window at the gaze pivot \cite{MRTKgazezoom,MRTKgazezoompage}. We extend this by a more detailed assessment of hand versus head/world-referenced UIs. There is essentially a trade-off:  As apparent in \autoref{nav}e, performing the zooming operation will also visibly reduce the size of the UI as it gets farther away. 

In contrast, in a head/world referenced UI (\autoref{nav}f), the hand operation occurs in an independent space to the UI, and can move in front or behind the UI. In both ways, the user operation includes clutching in case of multiple zoom operations; and  integration of a control-display ratio is relevant. However, this can highly depend on the application use case. For continuous zooming operations, more fine-grained translation functions must be implemented; for discrete operations (e.g., a folder structure with 3-5 levels), short hand motions can be sufficient.


\section{Applications}
We now explore design issues and opportunities for (1) system-wide application of the \system  concept in a concrete system implementation, and (2)  specific applications by way of technology  probes \cite{Hutchinson03}. An overview of the supported applications, and how each modality is utilised  is presented in \autoref{tab:applications}.

\begin{table}[]
\small
    \centering
    \caption{Applications in the \system system. All applications use gaze to point, as it is translated into (X,Y) UI coordinates, accompanied by a pinch-click gesture to select. Use of pinch-drag in one or more directions (X-horizontal, Y-vertical, Z-depth) is application-specific.}
    \begin{tabular}{lccc|c|ccc}
    \textbf{Application} & \multicolumn{3}{c}{\textbf{Gaze}}&\textbf{Pinch-click}&\multicolumn{3}{c}{\textbf{Pinch-drag}}\\
    &  X&Y&Z&&X&Y&Z\\
    \toprule
     Main menu & $\times$ & $\times$ && $\times$ &&&\\
    \toprule
     {Favorites Folder} & $\times$ & $\times$ && $\times$ &\\
     Notifications & $\times$ & $\times$ && $\times$ &&&\\
     {Music Player} & $\times$ & $\times$ && $\times$ &&&\\
     \midrule
     Downloads Folder & $\times$ & $\times$ && $\times$ & $\times$ &\\
     {Image Gallery} & $\times$ & $\times$ &&  &&& $\times$\\
     {Map Viewer} & $\times$ & $\times$ & & $\times$ & $\times$ &  $\times$& $\times$\\
    \end{tabular}
    \label{tab:applications}
\end{table}

\subsection{Implementation}
We used the Varjo XR-3\footnote{Varjo XR-3, \url{https://varjo.com/products/xr-3/}, accessed 04/02/2023} head-worn display with its integrated capabilities for pass-through AR, eye tracking  (200\,Hz, 1° reported accuracy) and Ultraleap hand tracking~\footnote{ Ultraleap Gemini, \url{https://www.ultraleap.com/tracking/gemini-hand-tracking-platform/}, accessed 04/02/2023}. The software was implemented in Unity.
The \textit{On-Hand} UI variant fixes the orientation in relation to the hand, and the UI is positioned just above the palm position of the monitored hand (by 4.5~cm). 
The UI may pick up on our natural hand motion and hand-tracking jitter. Smoothing the position helps, and through testing, we found that a parameter of 100 ms at a frame rate of 90Hz provided a good balance between UI stability and the delay in UI movement. For the \textit{Above-Hand} variant, given the hand-tracking quality, it is more susceptible to hand jitter as brief hand motions can cause large UI displacements. To counteract this effect, the UI ignores the hand rotation and  always faces the user regardless of its position. This allowed to flexibly move the UI in or out of the user's view through hand movement along all three axes while remaining relatively stable to be used regardless of the hand rotation. Lastly, a baseline \textit{Head-referenced} is included where the UI adapts its position to 55 cm in front of the forward vector of the headset user, oriented towards the user.
Item selection by gaze and pinch includes a default, hover (visual highlighting of icon background), and selected state. When buttons are close to each other, a single gaze point can theoretically fall into the empty space in between and void a selection intended by the user. To approach this, we assume if a gaze is within the UI bounds, a user is focusing on an interactive element, and hover/selection automatically snap to the closest target (not for non-interactive targets, e.g., when viewing images in the gallery).


\begin{figure}[t]\centering \includegraphics[width=1\columnwidth]{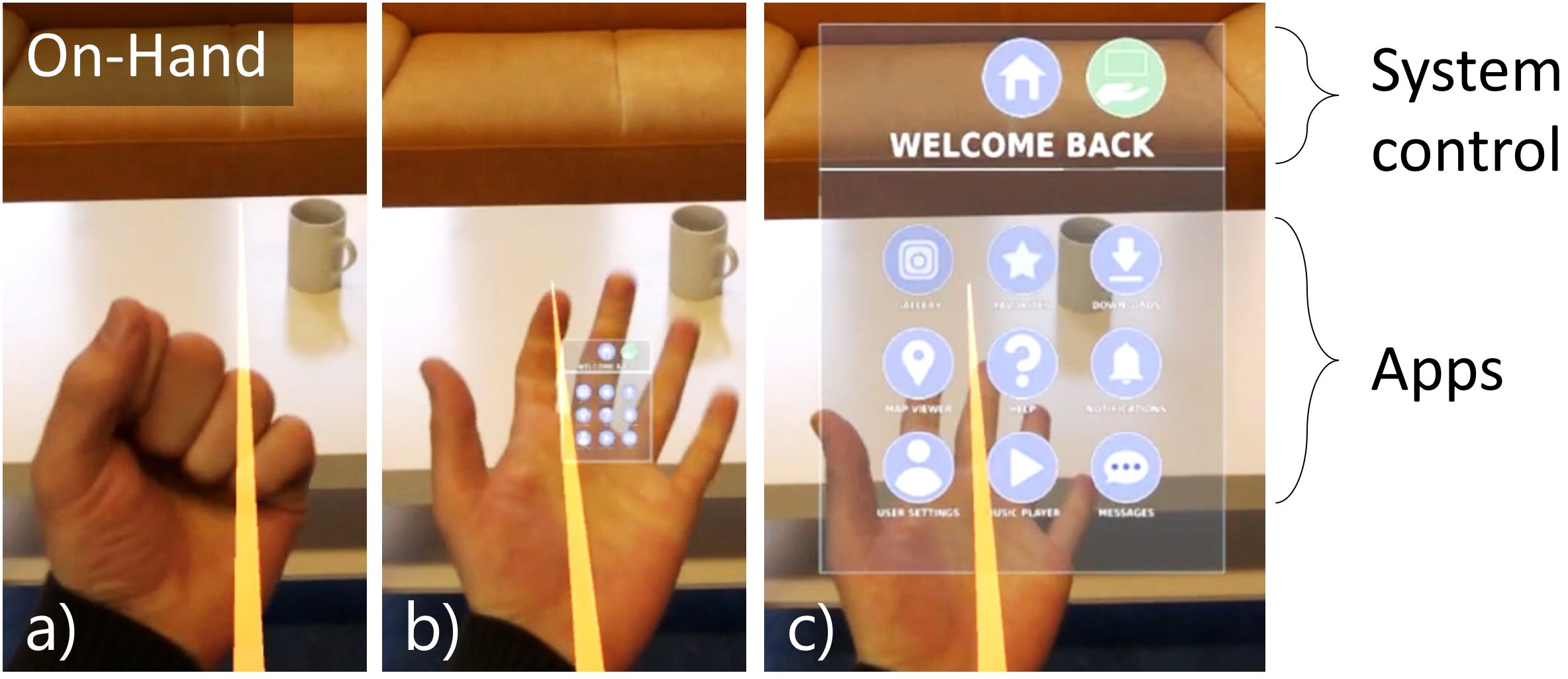} 
\vspace{-0.7cm}
\caption{
UI Activation supported  by a gradual scale-up summoning animation (a-c), to open the  home menu with a main  area and a top menu. The orange ray indicates the user's gaze.
} \label{activate}\end{figure}

\subsection{Home Menu}
The home menu allows to choose applications, which are then opened in individual UI windows.
It is opened by a gradual scale-up animation to provide a smooth transition (\autoref{activate}a-c). It offers a set of applications in the main UI area, and a fixed menu area at the top. The system remembers the last used application and will resume the state when the UI reactivates. 

\subsubsection{Implementational Details}
An additional top bar menu has been designed with a few buttons. The menu is active across applications, allowing the user to  return to the home menu on demand. For this, a dedicated button is reserved in the menu. A second button allows toggling the reference frame between \textit{On-Hand}, \textit{Above-Hand}, and \textit{Head-Referenced}, to offer users the choice (\autoref{activate}c-e). Setting this option will adopt the techniques for navigation to the respective reference frame  as elaborated in section 4.2.

\subsection{Music Player App}
The music player supports standard control buttons such as play, pause, and next, as well as a window of songs at the lower part of the UI window. It demonstrates the quickest to perform action in \system. To play a song, only a  'palm-open, look at song, and pinch' input sequence (assuming the music player is still open from last time of use) -- which, over time, appears like one cognitively unified interaction chunk, rather than a tedious sequence of operations. 

\subsection{Notifications  App}

\begin{figure}[t]\includegraphics[width=1\columnwidth]{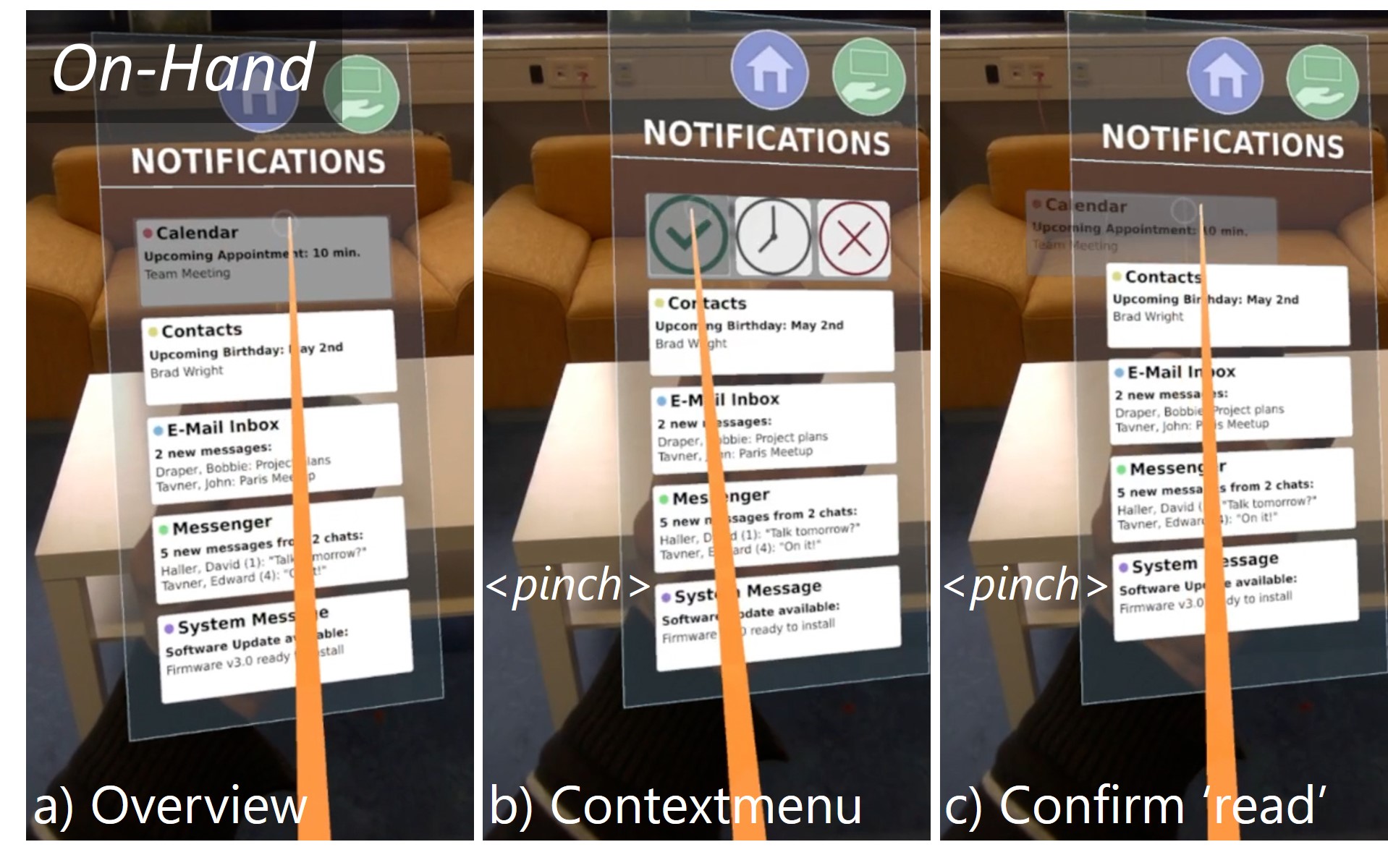} \centering
\vspace{-0.7cm}
\caption{
Notifications in a vertical list view (a) allow opening a context menu by selection (b) to respond (c).
}\label{notes}\end{figure}

The notification application provides an overview of notifications in a vertical list that users can quickly browse and respond to. The probe demonstrating in-page context menus that are integrated within each notification element (\autoref{notes}a). By default, a notification element displays the information on application, title, and description. At selection, each notification element transforms into a context menu with three selectable buttons to react to it (check, postpone, delete) (\autoref{notes}b-c).

\subsubsection{Implementational Details}
We tested persistent buttons next to the notifications as well, but found these to be needing too large spaces for robust selection and also they reduced the amount of information possible to be read on the notification UI element. Instead, we use the aforementioned two-selection model where the notification is opened first and then processed. 
An alternative to the default selection technique of the response to the notification would be to adopt shortcuts similar to touchscreen UIs. In the information view, the user can look at a notification and perform a pinch-drag-left gesture to immediately perform the 'check' reply (or a pinch-drag-right gesture for 'delete'). This variant provides extended expressive power, as multiple UI commands can be mapped to each logical dragging direction, similar to gaze and pinch based pie and marking menus \cite{Pfeuffer17}.

\begin{figure*}[h]\includegraphics[width=1\linewidth]{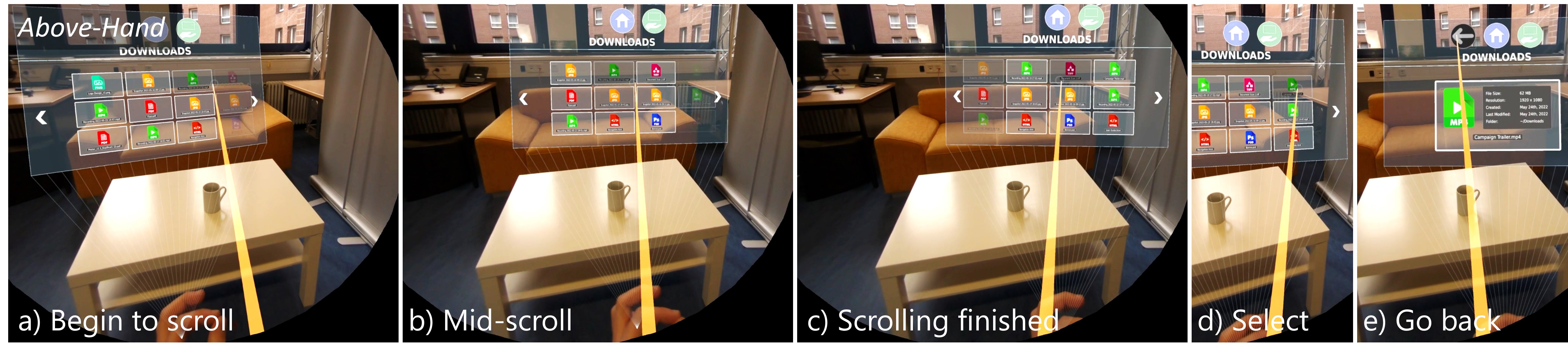} 
\vspace{-0.7cm}\caption{
The downloads app represents a grid menu, where pinch-drag gestures scroll the content (a-c), parallel to moving the UI as hand-attached. The user can immediately select a file (d) to view it (e), and return by selecting the back button (e-top).
}
\label{dls}\end{figure*}

\subsection{Downloads, Favorites}
Both Downloads and Favorites applications are standard 2D grid UIs that we started with as baselines. These demonstrate that the many potential 2D grid based applications can be easily supported. As an example, 'Downloads' includes a large number of files to demonstrate a horizontal scrollable grid view. 

\subsubsection{Scroll Interaction Technique}
An example of an Above-Hand scroll technique is shown in \autoref{dls}. Here, a user can browse through the downloaded files only by a pinch-drag gesture using a dynamic peephole view. To see more content on the right side of the grid, the UI is held at the left side of the field of view. Then, a pinch gesture is initiated from left (\autoref{dls}a) to the right (\autoref{dls}b-c). Afterwards, the user can select a file (\autoref{dls}d) and see details of it in the detail view. A back button, that appears in the detail view, is implemented to return to the grid (\autoref{dls}e).

\subsubsection{Implementational Details}
We developed the technique as described above, but also considered an alternative way of interaction that in principle comes with it. The user can  scroll without moving the hand, but by holding a pinch gesture and moving the own field of view (i.e., turning the head). Then, the UI scrolls equally as by hand.  In principle, both ways can be provided to users as they are complementary, and users get the choice.

\subsection{Image Gallery Application}
The Gallery  is a  grid UI where users can  scroll in vertical/horizontal directions (more pictures and folders), but also in-depth (overview, folder, picture) that we highlight in this probe. The application offers the user a method to rapidly view and traverse a large set of images in multiple hierarchy levels, with currently a depth or 3 layers supported.

\subsubsection{Deep-Gallery Interaction Technique}
\begin{figure}[t]\includegraphics[width=1\columnwidth]{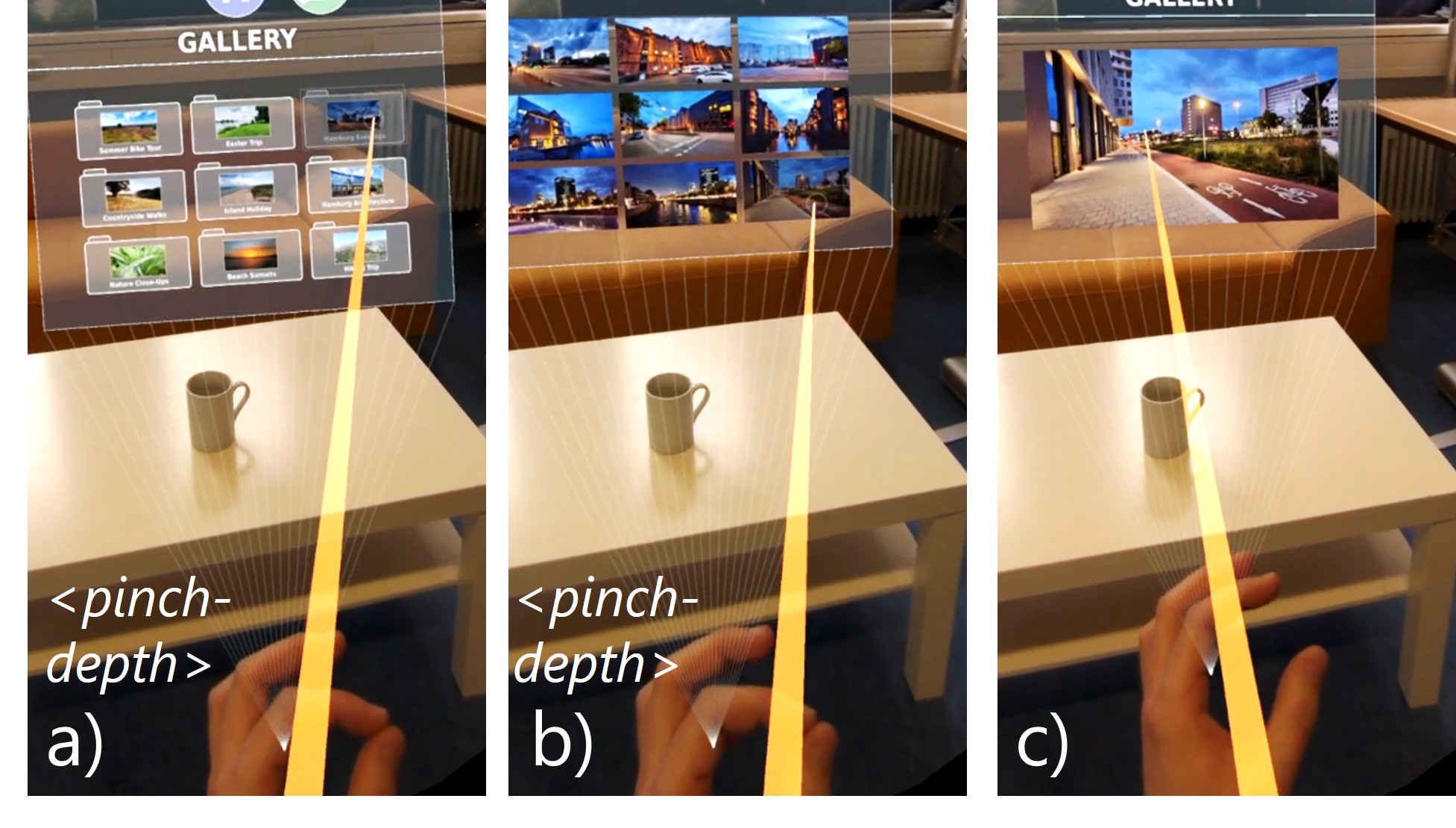} 
\vspace{-0.7cm}
\caption{
The Gallery involves a 3-level hierarchy traversed by gaze-selection and pinch for-/backward motion.
}\label{gal}\end{figure}

Navigating into depth can be performed in two ways. First, like the Downloads app, the user can select an item to traverse to the next depth level. The user performs a selection command each time. Second, a pinch-drag gesture can be performed in the for-/backward direction.  We virtually place the hierarchical layers 5~cm apart from each other and map the hand movement linearly to traverse these layers. This has an interesting appeal as it speaks to a metaphor of traversing depth levels, mapped  to depth motion of the hand.  The user can initiate a pinch gesture and hold it in the folder overview (\autoref{gal}a), then look at the folder and move the hand forward to open the image overview (\autoref{gal}b), and then look at an image and move the hand further forward to open the image (\autoref{gal}c).

\subsubsection{Implementational Details}
We tested various thresholds for the distance between 2 layers, and found that 5 cm provide a robust experience to still hold the hand in a layer, but also easily shift to the next while focusing gaze on the right direction. 
This highlights a trade-off between manual effort and hierarchical data traversal. For basic applications (e.g., Downloads), a single pinch-select gesture is sufficient and avoids UI repositioning, whereas with multiple hierarchy levels, using a single pinch-drag-depth technique can become beneficial.

\subsection{Navigation App for Map Interfaces}
Enlarging the area of visual attention can be a natural and accurate method for zooming in map navigation \cite{Klamka15,Pfeuffer14, Pfeuffer16par,Stellmach12}. 
In our \textit{Map Viewer}, users can spontaneously access geographical information and rapidly traverse the map. At begin, the prototype defaults to a full world map, at which the users can begin to navigate. The application probe supports sequential as well as simultaneous pan and zoom operations, providing the necessary set for navigation like on multi-touch based systems.

\subsubsection{Eye-Hand Pan-Zoom Interaction Technique}
\begin{figure}[t]\includegraphics[width=1\columnwidth]{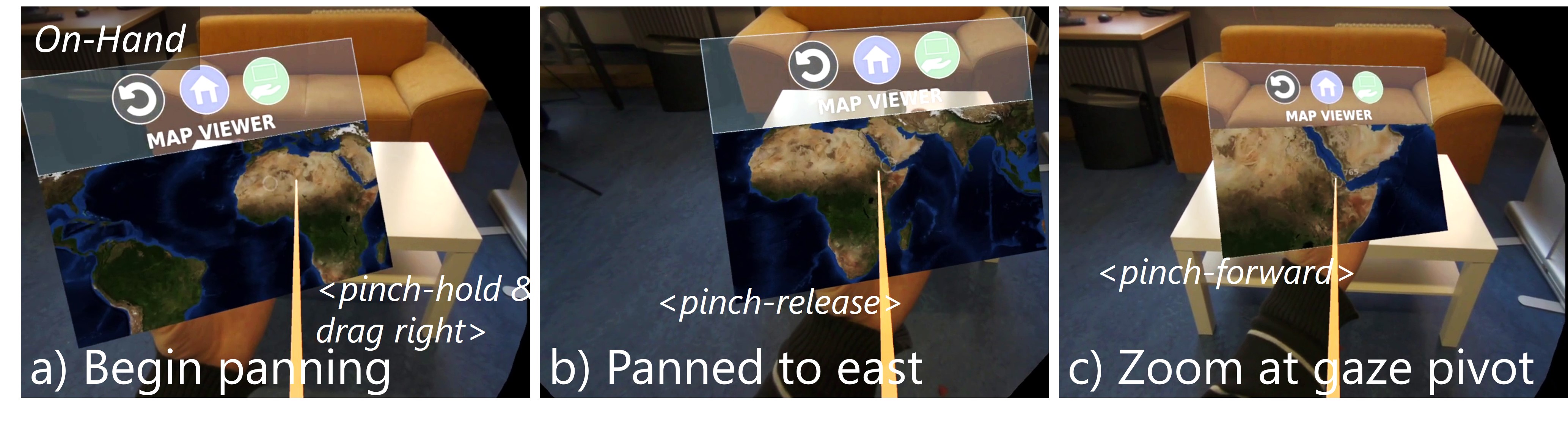} 
\vspace{-0.7cm}
\caption{
The Map Viewer integrates pan \& zoom  operated by a single pinch gesture and a gaze zooming pivot.
}\label{map}\end{figure}

Users can pan with a pinch-drag gesture without eye-tracking (\autoref{map}a-b). To zoom, users gaze at an area of interest and and can zoom into it by performing a pinch-drag toward from the UI, respectively (\autoref{map}c). 
Similar to the \textit{Downloads Folder}, the direction of panning is adjusted based on the placement of the menu: In a \textit{Head-referenced} mode, the content moves in the same direction as the hand, with the direction being reversed in the hand-referenced modes. To zoom in or out, depth navigation in form of continuous translation is used concurrently to 2D navigation via a dynamic peephole view. 

\subsubsection{Implementational Details}
Panning uses a control-display gain of 1:1, which is intuitive as if dragging the content (in the On Hand reference frame). Zooming transforms  hand motion to zoom in/out of the map with factor 2, i.e., more physical motion is applied to a zoom level of the image. We find that this allows to subsequently perform pinch dragging gestures of a length of approximately 5-20 cm to operate all typical pan and zoom operations. Zoom-out happens at pinch-drag backward, however instead of using the gaze pivot, the center of the map is used as the eyes are likely not indicative of the exact desired zoom-out position.

\section{Informal User Evaluation}
We conducted a qualitative usability evaluation, to explore the usability of the \system prototype,  in particular the usability of the different reference frames and interaction techniques.
Every user experienced the  \textit{On-Hand}, \textit{Above-Hand}, and  \textit{Head-referenced} placements. Each UI involved interactions with the Favorites Folder, Music Player, Image Gallery, Map Viewer applications.
 In each application, users received tasks to support a goal-directed experience,  displayed  on a virtual board. In the \textit{Favorites Folder} app, 
participants were given a filename and a property of that file (e.g., "Flyer.pdf", File Size property), and  were  instructed to find, view, and read out a file detail (file randomised, as in all tasks). In the  \textit{Music Player}, participants were given a track title to play. For the \textit{Gallery}, one image was overlaid with a three-digit number. Participants were given the album containing this image and instructed to find the image. In the \textit{Map Viewer}, the map was modified to contain three red circles, each marking a certain area on the map. Inside one of them, a three-digit number, displayed small enough to be only visible when zoomed in, was the search goal.

Users were first briefed,  filled out consent and demographic forms, and then began with the training. After showing and explaining the interactions, participants  freely trained until they felt confident and familiar enough to perform the tasks on-demand. Then, the task sessions began.
The active application and UI placement were set beforehand to ensure that the procedure was followed correctly. Users were free to take short breaks between apps. 
After each round, participants could provide written or verbal feedback. Users interacted with each application on average for 32.5 seconds (SD=7.4). Then, users were presented with the option to freely explore the system. Eight participants did so, spending between 3 and 28 minutes (M=14, Mdn=9). Lastly, an open-ended interview was conducted. The study lasted around 90 minutes on average. 

We recruited 18 paid participants (10 male, 8 female) via a university mailing list, messaging platforms, and personal contacts. All participants were right-handed and 20--40 years old (mean = 26.6, SD = 4.5). One of them wore glasses underneath the headset and three wore contact lenses. Participants rated their experience with different technologies on scales from 0 (no experience) to 4 (expert). The ratings were low to moderate for experience with VR/AR HMDs (M = 1.8, SD = 1.1), eye tracking UIs (M =  1.3, SD = 1.3), and mid-air gestures (M = 0.6, SD = 0.8).

\subsection{Results}

\system was overall well received, even though we encountered issues with more complex operations and hand-tracking constraints. 
 P15 stated ``\textit{It is very cool!}'' and P5 stated similarities to  smartphones: ``\textit{I can see how it can be used in the future  instead of smartphones or in addition to them.}''
The basic actions of UI activation and selection of elements  was received well by all users (described as '\textit{intuitive}', '\textit{easy}', '\textit{fun}', or '\textit{magical}'). Users found specific functions useful, such as changing the gaze selection during a longer pinch-drag interaction, either for quickly and continuously going through different images or for zooming in and out of different portions of the map by holding the pinch gesture while moving the hand back and forth. 


Pinch-drag interactions were called ``\textit{user-friendly}'' and  related  to traditional touchpad gestures, e.g.\ P7: ``\textit{A nice tool, and with a bit of adapting to it, it works quite well.}'' 
On the other hand, P10 stated: ``\textit{I like the idea, but I find it hard to control it how far you have to move the hand}'', while P7 pointed out that it needs getting used to and P16 called it ``\textit{very extraordinary in the beginning}''. This novelty aspect was also mentioned by P8, who stated: ``[I]\textit{never thought about an action like that, but once you know it, it works great}''. Meaning, the combined use of UI positioning and content alteration needs more getting accustomed to than the selection principle, before it worked well.

Gaze and pinch-drag-based depth navigation  was more positively received for the Gallery, which used only movement along one axis than for the Map Viewer. As stated by P1: ``\textit{The dragging back and forth was very smooth and exact. Mixed with the dragging to the left and right} [to see maps],\textit{ it was hard to handle.}''. In addition, P15 noted: ``\textit{I like it, particularly for the gallery. For the map app it was a bit more demanding to figure out how to do it properly.}''
Participants liked the idea of ``\textit{holding}'' the contents of the UI on their hand in the On-Hand condition. For example, P5 compared this to holding a smartphone and stated that they liked this placement concept the best, although they found it more difficult to use as of issues with tracking. Specifically for the Gallery app, the idea of holding the UI with an image On-Hand and being able to show it to others in a shared AR scenario was also mentioned by two participants as a positive, although they both added that it needs a more stable hand tracking. 
%
P1 and P7 reported for hand-attached UIs that small hand movements such as executing a pinch can shift the UI, possibly interfering with the user's gaze selection. 

The eyes can naturally follow the hand, which was received differently across placements. With the Head-Referenced and Above-Hand UIs, the eyes can lead to a hand-following behaviour which detracts from the main UI. E.g., \textit{P1} stated: ``\textit{When I move my hand, my first intuition is to look at my moving hand, but then I lose track of the menu and I have to re-orientate.}''.   For the On-Hand UI, P5 mentioned that the UI is positioned where their eyes wanted to look naturally and P14 pointed out that this also fits with panning the map by moving the hand with the UI on it, where the eyes can follow the hands. A potential reason is that users are familiar to this from smartphone use. 


\section{Discussion}
In this paper, we introduced \system as a unimanual eye-hand AR menu system. We described UI concept design, interaction techniques, example applications and an informal user study.
%

\subsection{Design Insights and Recommendations}
Based on the design and prototyping experience and the findings from the evaluation, we formulate  the main insights and recommendations from this research.

\textbf{Simultaneous and Interchangeable Eye-Hand Interaction.}
To point at a button, users may either glance at it or move the UI (and hence the button) by hand into the gaze point. This interchangeability of modalities is extremely important for providing users with more alternatives, balancing the interaction burden between modalities, accommodating possible gaze error, and supporting apps with small target.

In a non-handheld UI, eye tracking accuracy frequently needs careful target size considerations and may be user-specific.  A portable UI, on the other hand, allows the user to achieve the appropriate visual target size because bringing the target closer increases size implicitly. Adapting UI orientation to the hand, similar to smartphones, appears natural at first glance. However, because rotating the UI targets reduces visual target size, it has a negative influence on ocular input quality. Auto rotation towards the user is a practical solution to this problem. Overall, by making accuracy an implicit user-orchestrated approach, it simplifies interacting with the eyes and hands.

\textbf{Activation and Interaction Task Allocation.} Tracking hand gestures beyond a standard pinch gesture becomes increasingly difficult for the hand tracking sensor. The combination of palm open to activate and eye-hand input to interact while keeping the hand open provided a practical trade-off between interaction expressiveness and  tracking accuracy.
While we first assumed it would be utilised mostly when the palm is looking up or towards the user, it was frequently used when the palm was facing away from the user. This enabled for superior pinch tracking because the two fingers were more visible, and it was also more ergonomic interaction.

When the UI is active, supporting all fundamental 2D interactions (selection, scrolling) allows for rapid activities on mobile (changing songs, checking notifications, and so on). The 3D interaction tasks presented us with a different image. Users found one-directional and basic actions, such as discrete navigation in depth through a pinch in depth direction, to be usable, such as in our gallery app, where users leap to the next hierarchical level. More degrees of freedom, such as integral, continuous 3d pan and zoom in map navigation, are not advised, as coordination of the degrees of freedom while holding the UI overwhelmed the users.

\textbf{UI Navigation.} Most of our UI methods have shown to be effective in providing expressive power comparable to one-handed interaction with a smartphone. A top menu home button lets users to utilise gaze and pinch to swiftly transition from one app to the home menu to a new app, making OS-level navigation effortless. In order to navigate within the app, pinching and dragging (without eye-tracking) is appropriate for scrolling, panning, and swiping. 

A key interaction problem is that users may use hand movement to position the UI as well as manipulate UI information. Standard world-referenced input mappings are not suitable by design-- for which we implemented a peephole metaphor \cite{Fitzmaurice93,Mehra06}. We did not find any issues with its usability and users were generally positive about the various scrolling and panning tasks, thus recommend its use to establish an intuitive navigation experience.

\textbf{Reference Frame Trade-Offs.}  The On-hand frame necessitates the user either moving the UI into view or moving the FoV to the hand, resulting in physical weariness. However, because the UI is directly associated with the hand in space, it has  a stabilising effect to the UI in a position. In contrast, the Above-Hand UI, in theory, addresses the previous issue since users may drop their hand to a convenient position while the UI is in a comfortable view location. We found, however, that it is less stable since little hand jitter translates to large UI motion effects.
Adding system choices that allow users to switch between reference frames to obtain the best of all worlds depending on the scenario can be suitable, e.g., a quick 'lift-up' gesture to 'throw' the UI from On-Hand to Above-Hand, or even employing adaptation algorithms that takes the user's reachability and visibility constraints into account \cite{Belo22}.

\subsection{Limitations and Future Work}
In this work we did not conduct a formal evaluation to evaluate performance and better understand strength and weaknesses. How long, for example, does it take to discover and open an app  while on the go, in contrast to existing AR UIs? How may the interaction technique, as well as people and objects around the user potentially distract the user and affect cognitive load? In principle, both eye-hand and hands-only control can be supported complementarily, -- how does switching input contexts affect the usability? 
Further in this work we considered one-handed interaction mainly for tasks of pointing and navigation. However, other tasks such as text entry \cite{Lystbaek22text} can be as relevant and it is open how a one-handed way can be designed.

\section{Conclusion}
In this work, we designed, implemented, and evaluated \system, a new one-handed hand-attached UI for spontaneous interaction in everyday AR. We demonstrated this through a system prototype that offers input capabilities similar to modern smartphones  -- to select, scroll, navigate, and interact with mobile apps fluidly with one hand. We described applications in detail, highlighting individual design issues and interaction techniques. Our informal evaluation provided important insights on the usability and applicability of the concept. The limits in the user's expressiveness are indicated in more complex operations, where users must coordinate multiple dimensions in parallel, indicating room for improvements or potentially the border to employ hands-only. But the fundamental interaction principle is promising as a diverse, easy-to-use interaction style, facilitating quick actions through a spontaneously activated UI. As limitations, the potential of several interaction concepts was hampered by the hand tracking quality (e.g., when holding and interacting in parallel), which could be potentially revisited in future in case the sensors improve. 

In contrast to hand-based interaction research, the work for eye-hand interaction is at its infancy. With current industrial devices becoming just ready for a novel eye-hand interaction paradigm for consumers, it becomes ever more important to explore the vast design space of eye-tracking advancements in all those contexts that have been reserved for hands only. Our work expands the research line for gaze and hand based UI systems \cite{Pfeuffer14,Pfeuffer16,Pfeuffer17}, and informs the design of multi-modal, flexibly-placed, on-demand AR UIs for a future where our mobile life spans far beyond holding a mobile device every time we want to interact, toward non-invasive but always-available UIs to seamlessly engage with digital content on the go. 

\bibliographystyle{ACM-Reference-Format}
\bibliography{refs}


\end{document}